\RequirePackage{fix-cm}
\documentclass[smallextended]{svjour3}       

\smartqed  
\usepackage{graphicx}
\usepackage{amsfonts}
\newcounter{icirc}
\usepackage{tikz}
\usetikzlibrary{shapes,arrows}
\usetikzlibrary{intersections}
\usepackage{braket}
\usepackage{physics}
\usepackage{youngtab}
\usepackage{bbm}
\usepackage{comment}
\usepackage[T1]{fontenc} 
\def\CL{{\mathcal L}}
\def\q{{\mathbbm{q}}}
\newcommand{\Z}{{\mathbb Z}}
\newcommand{\C}{{\mathbb C}}
\def\CI{{\mathcal I}}
\def\ie{{\textit{i.e.}}}
\newcommand{\Q}{{\mathbb Q}}
\def\lk{{\ell k}}
\def\CW{{\mathcal W}}
\def\Tor{{\mathrm{Tor}\,}}

\def\Tor{{\mathrm{Tor}\,}}

\newcommand{\be}{\begin{equation}}
	\newcommand{\ee}{\end{equation}}
\def\CM{{\mathcal M}}

\def\CM{{\mathcal M}}

\begin{document}

\title{$\hat Z$- invariant for $SO(3)$ and $OSp(1|2)$ Groups
}


\author{Sachin Chauhan         \and
        Pichai Ramadevi 
}


\institute{Sachin Chauhan \at
              Department of Physics, Indian Institute of Technology Bombay\\Powai, Mumbai, 400076, India\\
              \email{sachinchauhan@iitb.ac.in}           
           \and
           Pichai Ramadevi \at
              Department of Physics, Indian Institute of Technology Bombay\\Powai, Mumbai, 400076, India\\
              \email{ramadevi@iitb.ac.in}
}

\date{Received: date / Accepted: date}

\maketitle

\begin{abstract}
Three-manifold invariants  $\hat Z$ (``$Z$-hat''), also  known as homological blocks, are $q$-series with integer coefficients. Explicit $q$-series  form for $\hat Z$ is known for $SU(2)$ group, supergroup $SU(2|1)$ and ortho-symplectic supergroup $OSp(2|2)$. We focus on $\hat Z$ for  $SO(3)$ group and orthosymplectic supergroup $OSp(1|2)$ in this paper. Particularly, the change of variable relating $SU(2)$ link invariants to the $SO(3)$ \& $OSp(1|2)$ link invariants plays a crucial role in explicitly writing the $q$-series.
\end{abstract}

\section{Introduction}
\label{sec:intro}
Knot theory has attracted attention from both mathematicians and physicists during the last 40 years. The seminal work of Witten\cite{Witten:1988hf} giving a three-dimensional definition for Jones polynomials of knots and links, using $SU(2)$ Chern-Simons theory on $S^3$,  triggered  a tower of new colored link invariants. Such new invariants are given by expectation value of  Wilson loops  carrying higher dimensional representation $R \in \mathcal G$  in Chern-Simons theory where $\mathcal G$ denotes gauge group. These link invariants  are in variable $\q$ which depends on the rank of the gauge group $\mathcal G$ and the Chern-Simons coupling constant $k \in \mathbb Z$ \Big(For eg: when $\mathcal G=SU(N)$ then $\q=\text{exp}\left(\frac{2\pi i}{k+N}\right)$\Big). Witten's approach also gives three-manifold invariant $Z_k^{\mathcal G} [M;\q]$, called Chern-Simons partition function for manifold $M$, obtained from surgery of framed links on $S^3$(Lickorish-Wallace theorem\cite{10.2307/1970373,wallace_1960}). Witten-Reshitikhin-Turaev (WRT)  invariants $\tau_k^{\mathcal G}[M;\q]$ known in the  mathematics literature are proportional to the Chern-Simons partition function:

\begin{equation}
	Z_k^{\mathcal G}[M;\q] = {\tau_k^{\mathcal G}[M;\q] \over \tau_k^\mathcal G[S^2 \times S^1;\q]}~.\label{wrtdef}
\end{equation}

These WRT invariants can be written in terms of the colored invariants of framed links\cite{10.2307/1970373,wallace_1960,Kaul:2000xe,Ramadevi:1999nd}.

It was puzzling observation that the colored knot polynomials appear as  Laurent series with integer coefficients. There must be an underlying topological interpretation of such integer coefficients. This question was answered both from mathematics and physics perspective.  Initial work of  Khovanov\cite{khovanov2000categorification} titled `cateforification'   followed by other papers on bi-graded homology theory including Khovanov-Rozansky homology led to new homological invariants. Thus the integer coefficients of the colored knot polynomials are interpreted as the dimensions of vector space of homological theory. From topological strings and intersecting branes\cite{Ooguri:1999bv,Gopakumar:1998ii,Gopakumar:1998jq}, the integers of HOMFLY-PT polynomials are interpretable as counting of BPS states.  Further the  connections to knot homologies within topological string context was initiated in \cite{Gukov:2004hz} resulting in concrete predictions of homological invariants for some knots (see review \cite{Nawata:2015wya} and references therein). Such a  physics approach involving brane set up in $M$-theory\cite{Gukov:2017kmk,Gukov:2016gkn,Mikhaylov:2014aoa,Ferrari:2020avq} suggests the plausibility of categorification of WRT invariants $\tau_k^{\mathcal G}[M;\q]$ for three-manifolds.  However,  the WRT invariants for simple three-manifolds are not a Laurent series with integer coefficients.   

The detailed discussion on $U(N)$ Chern-Simons partition function on Lens space $L(p,1)\equiv S^3/\mathbb Z_p$ (see section 6 of \cite{Gukov:2016gkn}) shows a  basis transformation $Z_k^{\mathcal G}[M;\q] \underrightarrow{~~\mathcal S~~} {\hat Z}^{\mathcal G}[M;q]$ so that  $\hat Z$  are $q$-series(where variable $q$ is an arbitrary complex number inside a unit disk) with integer coefficients (GPPV conjecture\cite{Gukov:2017kmk}). These $\hat Z$  are called the homological blocks of WRT invariants of three-manifolds $M$. Physically, the new three-manifold invariants  $\hat Z^{\mathcal G}[M;q]$ is the partition function $Z_{T^{\mathcal G}[M]}[D^2 \times S^1]$ for simple Lie groups. Here $T^{\mathcal G}[M]$ denote the effective 3d $\mathcal N=2$ theory on $D^2 \times S^1$ obtained by reducing 6d $(2,0)$ theory (describing dynamics of coincident $M5$ branes) on $M$. 

For a class of negative definite plumbed three-manifolds as well as link complements \cite{Gukov:2017kmk,Gukov:2019mnk,park2020higher,Chung:2018rea}, $\hat Z^{SU(N)}$ has been calculated.  Further, $\hat Z$ invariants for super unitary group $SU(n\vert m)$ supergroup with explicit $q$-series for  $SU(2\vert 1)$ is presented in \cite{Ferrari:2020avq}. Generalisation to  orthosymplectic supergroup $OSp(2|2n)$ with explicit $q$-series for $Osp(2|2)$\cite{chae2021towards} motivates us to look at $\hat Z$ for other gauge groups.

Our goal in this paper is to extract $\hat Z$ for the simplest orthogonal group $SO(3)$ and the simplest odd orthosymplectic supergroup $OSp(1|2)$.  We take the route of relating $SU(2)$ colored link invariants to the link invariants for these  two groups to obtain $\hat Z$ invariants. 

The plan of the paper is organised as follows. In section 2, we will review the developments on the invariants of knots, links and three-manifolds. We will first briefly present Chern-Simons theory and colored link invariants with explicit results for $SU(2)$ gauge group and indicate how colored $SO(3)$ and $OSp(1|2)$ link invariants can be obtained from the colored $SU(2)$ polynomials. Then, we will summarise the developments of the homological invariants. In section \ref{reviewsu2}, we  briefly  review 
$\hat Z$-series invariant for $SU(2)$ group for the negative definite plumbed three-manifolds. This will serve as a warmup to extend  to $SO(3)$ and $OSp(1|2)$ group which we will present in section \ref{zhatnew}. We summarize  the results and also indicate future directions to pursue in the concluding section \ref{conclusion}.

\section{Knots, Links and Three-manifold Invariants}
In this section, we will briefly summarise new invariants in knot theory from the physics approach as well as from the mathematics approach.
\subsection {Chern-Simons Field Theory Invariants}
Chern-Simons theory based on gauge group $\mathcal G$ is a Schwarz type topological field theory which provides a natural framework for study of knots, links and three-manifolds $M$. Chern-Simons action $S_{CS}^{\mathcal{G}}(A)$ is explicitly metric independent:
\begin{equation}
	S_{CS}^{\mathcal{G}}(A)=\frac{k}{4\pi}\int_M Tr\left(A\wedge dA+\frac{2}{3}A\wedge A\wedge A\right)~.
\end{equation}
Here $A$ is the matrix valued gauge connection based on gauge group $\mathcal G$ and $k\in \mathbb{Z}$ is the coupling constant. In fact, the expectation value of Wilson loop operators associated with any $m$-component link $\mathcal L_m$ are the  the link invariants:
\begin{multline} 
	V_{R_1,R_2, \ldots R_m}^{\mathcal G} [\mathcal L_m;\q]= \langle W_{R_1,R_2,\ldots R_m}[\mathcal L_m]\rangle= 
	{\int {\mathcal D}A \exp(iS_{CS})  \overbrace{P\left(\prod_i  \Tr_{R_i} exp \oint_{\mathcal K_i} A\right)}^{W_{R_1,R_2,\ldots R_m}[\mathcal L_m]} \over {\underbrace{\int {\mathcal D} A \exp(iS_{CS})}_{Z^{\mathcal G}_k[M;\q]}}}~,
\end{multline}
where $\mathcal K_i$'s denote the component knots of link $\mathcal L_m$ carrying representations $R_i$'s of gauge group $\mathcal G$ and $Z^{\mathcal G}_k[M;\q]$ defines the Chern-Simons partition function encoding the topology of the three-manifold $M$.

Exploiting the connection between Chern-Simons theory, based on  group $\mathcal G$, and the corresponding Wess-Zumino-Witten (WZW) conformal field theory with the affine Lie algebra symmetry $\mathfrak g_k$,  the  invariants of these links embedded in a three-sphere $M=S^3$ can be explicitly written in variable $\q$ :
\begin{equation} 
	\q= \exp({2\pi i \over k+C_v})~,
\end{equation}
which depends on the coupling constant $k$ and the dual Coxeter number $C_v$ of the group $\mathcal G$. These link invariants  include the well-known polynomials in the knot theory literature. 

\begin{center}
	\begin{tabular}{|c|c|c|c|} \hline
		$\mathcal G$&$R$&Invariant\\
		\hline
		&&\\
		$SU(2)$ &$\yng(1)$ & Jones \\
		$SU(N)$ &$\yng(1)$ & HOMFLY-PT \\
		$SO(N)$ & defining & Kauffman \\
		\hline
	\end{tabular}
\end{center}
\subsubsection{Link Invariants}
\label{sec2.1.1}
\noindent
As indicated in the above table, Jones' polynomial corresponds to the fundamental representation $R=\tiny {\yng(1)}\equiv 1 \in SU(2)$ placed on all the component knots:
\begin{equation}
	V_{1,1,1,\ldots 1}^{SU(2)} [\mathcal L_m;\q] \equiv J_{2,2,\ldots 2}\left[\mathcal L_m; \q=\exp({2\pi i \over k+2})\right]~,
\end{equation}
where the subscript `2'  in Jones  polynomial $J_{2,2,2,\ldots}[\mathcal L_m;\q]$ denotes the dimension of $R={\tiny {\yng(1)}}$. Higher dimensional representations   placed on the component knots $R_i =\underbrace{\tiny {\yng(4)}}_{n_i-1}\equiv n_i-1\in SU(2)$  are the colored Jones invariants:
\begin{equation}
	V_{n_1-1,n_2-1,n_3-1,\ldots n_m-1}^{SU(2)} [\mathcal L_m;\q] \equiv J_{n_1,n_2,\ldots n_m}\left[\mathcal L_m; \q=\exp({2\pi i \over k+2})\right]~,
\end{equation} 
and the invariants with these representations belonging to $SU(N)$ ($(SO(N)$)  are known as colored HOMFLY-PT (colored Kauffman)  invariants. For clarity, we will restrict to $SU(2)$ group to write the invariants explicitly in terms of $\q$ variable. 

We work with the following unknot ($\bigcirc$) normalisation:
\begin{equation} 
	J_{n+1} [ \bigcirc;\q] = {\rm dim}_\q \underbrace{\yng(4)}_n={ \q^{(n+1)\over 2 } - \q^{-{(n+1) \over 2}} \over \q^{1\over 2} - \q^{-{1\over 2}}}= {\sin({ \pi (n+1) \over k+2})\over \sin({ \pi  \over k+2})}=
	{S_{0 n}\over S_{00}},
\end{equation}
where ${\rm dim}_\q \underbrace{\yng(3)}_n $ denotes quantum dimension of the representation $\underbrace{\yng(3)}_n$ and $S_{n_1 n_2}$ are the modular transformation matrix elements of the $\mathfrak {su}(2)_k$ WZW conformal field theory whose action on the characters is $\chi_{n_1} (\tau)  ~~\underrightarrow{~~~S~~~}~~ \chi_{n_2} \left(-{1\over \tau}\right)~,$ where $\tau$ denotes the modular parameter.  These knot and link polynomials with  the above unknot normalisation are referred as unnormalised colored Jones invariant.

\begin{figure}[ht]
	\centering
	\includegraphics[scale=0.2]{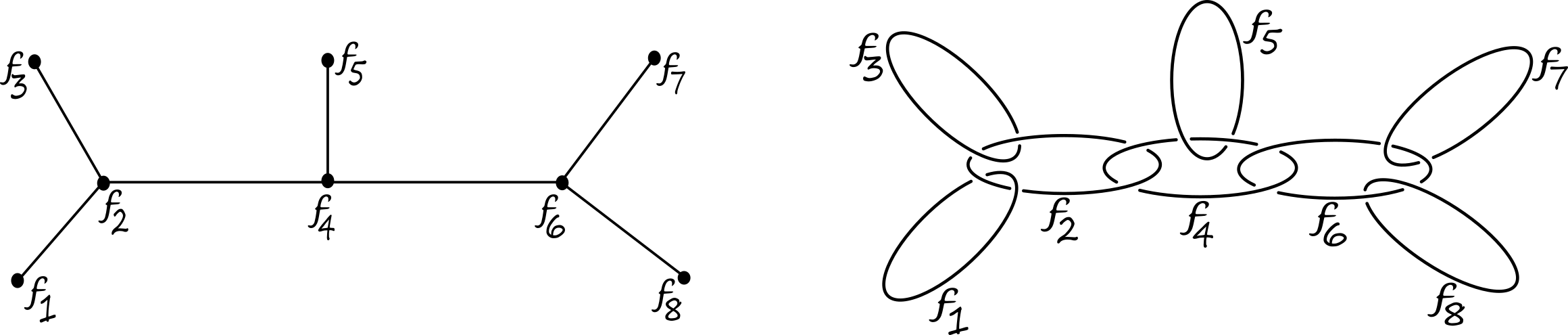}
	\caption{An example of a plumbing graph $\Gamma$ (left) and the corresponding link $\CL(\Gamma)$ of framed unknots in $S^3$ (right). 
	}
	\label{fig:plumbing-example}
\end{figure}

For framed unknots with framing number $f$, the invariant will be 
\begin{equation} 
	J_{n+1} [ \bigcirc_f;\q] = \q^{f [{(n+1)^2-1 \over 4}]} { \q^{(n+1)\over 2 } - \q^{-{(n+1) \over 2}} \over \q^{1\over 2} - \q^{-{1\over 2}}}\propto (T_{nn})^f	{S_{0 n}\over S_{00}},
\end{equation}
where the action of the modular transformation matrix $T$ on characters is \\$ \chi_n(\tau) ~~\underrightarrow{~~~T~~~~} \chi_n(\tau+1)~.$
The colored Jones invariant for the Hopf link  can also be written in terms of $S$ matrix:
\begin{equation} 
	J_{n_1+1, n_2 +1} [H;\q] =  \left({\q^{\frac{(n_1+1)(n_2+1)}{2}} - \q^{-\frac{(n_1+1)(n_2 +1)}{2}}\over \q^{\frac{1}{2}} - \q^{-\frac{1}{2}}}\right)= 
	{S_{n_1 n_2}\over S_{00}}.
\end{equation}
The invariant for a  framed Hopf link $H(f_1,f_2)$, with framing numbers $f_1$ and $f_2$ on the two component knots, in terms of $T$ and $S$ matrices is 
\begin{equation} 
	J_{n_1+1, n_2 +1} [H(f_1,f_2);\q] \propto   (T_{n_1 n_1})^{f_1} (T_{n_2 n_2})^{f_2}	{S_{n_1 n_2}\over S_{00}}~ \label {hopf}~.
\end{equation}
We will look at a class of links obtained as a connected sum of framed Hopf links. For instance,  the invariant for the connected sum of two framed Hopf links $H(f_1,f_2) \#H(f_2,f_3)$ will be
\begin{eqnarray}
	J_{n_1+1, n_2 +1,n_3+1} [H(f_1,f_2)\# H(f_2,f_3);\q]& \propto&
	{ \prod_{i=1}^3T_{n_i n_i}^{f_i}} {S_{n_1 n_2}\over S_{00}} {S_{n_2 n_3}\over S_{n_2 0}} \label {conn}\\
	& =& {\prod_{i=1}^2J_{n_i+1, n_{i+1} +1} [H(f_i,f_{i+1});\q]  \over J_{n_2+1} [\bigcirc;\q]}~.\nonumber~
\end{eqnarray}
Such a connected sum of two framed  Hopf links, which is a 3-component link,  can be denoted as a  linear  graph \\$${\small  
	\begin{tikzpicture}[circ/.style={circle,draw,inner sep=2pt,label=above:$#1$,
			/utils/exec=\stepcounter{icirc},name=c-\number\value{icirc},
			node contents={}},scale=1.5]
		\path (-2.1,0) node[circ=f_1] (-2,0) coordinate (p-1) (-1,0) node[circ=f_2] 
		(0,0) node[circ=f_3];
		\draw foreach \X in {1} {(c-\X) -- (p-\X) (c-\the\numexpr\X+1) -- (c-3)};
		\draw foreach \X in {1} {(c-\the\numexpr\X+1) -- (p-\X)};
	\end{tikzpicture}
}$$
with three vertices labeled by the framing numbers and the edges connecting the adjacent vertices. These are known as `plumbing graphs'.  Another  plumbing graph $\Gamma$  with 8 vertices denoting the link $\mathcal L(\Gamma)$  (the connected sum of many framed Hopf links)  is illustrated in Figure \ref{fig:plumbing-example}. The colored  invariant for these links $\mathcal L(\Gamma)$ can be written  in terms of $S$ and $T$ matrices.

For a general  $m$ vertex plumbing graph with vertices $v_1,v_2, \ldots v_m \in V$  labelled by framing numbers $f_1,f_2, \ldots f_m$,  there can be one or more edges connecting a vertex $v$ with the other vertices. The degree of any vertex $v$ ($\text{deg}(v)$) is equal to the total number of edges intersecting $v$.  For the graph in Figure 1,   $\text{deg}(2)=\text{deg}(4)=\text{deg}(6)=3$. The colored Jones' invariant for any plumbing graph $\Gamma$ is
\begin{equation} 
	J_{n_1+1, n_2+1, \ldots n_m+1}[\mathcal L;\q] \propto {1 \over S_{00}} \prod_{i=1}^m \{(T_{n_i n_i})^{f_i} (S_{0 n_i})^{1-\text{deg}(v_i)}\}
	\prod_{(v_1,v_2)\;\in\;\text{Edges} } (S_{n_{v_1} n_{v_2}})~. \label {conn1}
\end{equation}
Even though we have presented the colored Jones
invariants (\ref{hopf}, \ref{conn}, \ref{conn1}),  the formal expression of these link invariants in terms of $S$ and $T$ matrices are applicable for any arbitrary gauge group $\mathcal G$. 

$\bullet$ \underline{\bf{$SO(3)$ and $OSp(1|2)$ Link invariants}}

\noindent
Using group theory arguments, it is possible to relate colored link invariants between different groups. For instance,  the representations of the $SO(3)$  can be identified with a subset of  $SU(2)$ representations. As a consequence, the $SO(3)$ link invariants  can be related to the colored Jones invariants as follows:
\begin{equation}
	V_{n_1,n_2,n_3,\ldots n_m}^{SO(3)} \left[\mathcal L_m;Q=\exp({2\pi i \over K+1})\right] = J_{2n_1+1,2n_2+1,\ldots 2n_m+1}[\mathcal L_m; \q]{\big\vert}_{\q^2=Q}~,\label{so3}
\end{equation}
where the level $K$ of the affine $\mathfrak {so}(3)_K$  Lie algebra must be an even integer( $K\in 2 \mathbb Z$). 

Similarly, the  representations of the orthosymplectic supergroup $OSp(1|2)$ can be related to the representations of the $SU(2)$ group  from the  study of $\mathfrak{osp}(1|2)_{\hat K}$ WZW conformal field theory  and the link invariants \\
$V_{n_1,n_2,n_3,\ldots n_m}^{OSp(1|2)} \left[\mathcal L_m;\hat Q=\exp({2\pi i \over 2{\hat K}+3})\right]$ 
\cite{Ennes:1997kx}.  
Particularly, there is a precise identification of the polynomial variable $\hat Q$ to $SU(2)$ variable $\q$. Further, the fusion rules of the primary fields of  $\mathfrak{osp}(1|2)_{\hat K}$ WZW conformal field theory can be compared to integer spin primary fields of the $\mathfrak{su}(2)_k$.  Particularly, the $\hat S$  and $\hat T$-matrices of $\mathfrak{osp}(1|2)_{\hat K}$ :
\begin{eqnarray}
\hat S_{n_1 n_2} &=& \sqrt{4 \over 2\hat K+3} (-1)^{n_1+n_2} \cos\left[{ (2n_1+1) (2n_2+1) \over 2(2\hat K+3)} \pi \right]~~,\\
\hat T_{n_1,n_2}&\propto&  \delta_{n_1,n_2} {\hat Q}^{[\frac{(2n_1+1)^2-1]}{4}}~~~,
\end{eqnarray}
are related  to the $S$ and $T$ matrices of  $\mathfrak{su}(2)_k$ in the following way:
\begin{equation} 
\hat S_{n_1,n_2} = S_{2n_1,2n_2}{\big\vert}_{\q = -\hat Q}~;~
\hat T_{n_1,n_1} = T_{2n_1,2n_1}{\big\vert}_{\q = -\hat Q}
\end{equation}
Using these relations, we can show that the   $OSp(1|2)$ colored invariant match the 
 colored Jones invariant for any arbitrary link $\mathcal L_m$ in the following way:
\begin{equation}
	V_{n_1,n_2,n_3,\ldots n_m}^{OSp(1|2)} \left[\mathcal L_m;\hat Q=\exp({2\pi i \over 2{\hat K}+3})\right] = \epsilon J_{2n_1+1,2n_2+1,\ldots 2n_m+1}[\mathcal L_m; \q]{\big\vert}_{\q=-\hat Q}~,\label{osp}
\end{equation}
where $\epsilon$ could be $\pm 1$ depending on the link $\mathcal L$ and the representations $n_i$'s.  For example, the colored $OSp(1|2)$ invariant for framed Hopf link is 
\begin{eqnarray} 
	V_{n_1, n_2 } ^{OSp(1|2)} [H(f_1,f_2);\hat Q] &=& 
	{\hat Q}^{\frac{f_1((2n_1+1)^2-1)}{4}}  {\hat Q}^{\frac{f_2((2n_1+1)^2-1)}{4}}(-1)^{(n_1+n_2)}\times\nonumber\\
	~&~&\left({{\hat Q}^{\frac{(2n_1+1)(2n_2+1)}{2}} + {\hat Q}^{-\frac{(2n_1+1)(2n_2 +1)}{2}}  \over {\hat Q}^{\frac{1}{2}} +{\hat Q}^{-\frac{1}{2}}}\right)\\
	&=& J_{2n_1+1, 2n_2+1} [H(f_1,f_2),-\hat Q]~.\nonumber
\end{eqnarray}
In fact, for any link $\mathcal L(\Gamma)$ denoted  by the graph $\Gamma$, the invariants will be 
\begin{eqnarray}
	V_{n_1, n_2,\ldots n_m } ^{OSp(1|2)} [\mathcal L(\Gamma);\hat Q] &=& {1  \over {\hat Q}^{\frac{1}{2}} +{\hat Q}^{-\frac{1}{2}}}\prod_{i=1} ^m (-1)^{n_i}
	{\hat Q}^{\frac{f_i((2n_i+1)^2-1)}{4}}\nonumber\\
	~&~&\left( {\hat Q}^{\frac{2n_i+1}{2} } + {\hat Q}^{-{\frac{2n_i+1}{2}}} \right)^{\text{deg} (v_i) -1}\nonumber\\
	~&~&\prod_{(v_1,v_2)\;\in\; {\rm Edges} } {\left({\hat Q}^{\frac{(2n_{v_1}+1)(2n_{v_2}+1)}{2}} + {\hat Q}^{-\frac{(2n_{v_1}+1)(2n_{v_2} +1)}{2}}\right)}~.
	\label{eqn2.15}
\end{eqnarray}
As three-manifolds can be constructed by a surgery procedure on any framed link, 
the Chern-Simons partition function/WRT invariant (\ref{wrtdef}) can be written in terms of link invariants\cite{10.2307/1970373,wallace_1960,Kaul:2000xe,Ramadevi:1999nd}. We will now present the salient features of such WRT invariants.
\subsubsection{Three-Manifold Invariants}
\label{sec2.1.2}
\begin{figure}[ht]
	\centering
	\includegraphics[width=12cm, height=3cm]{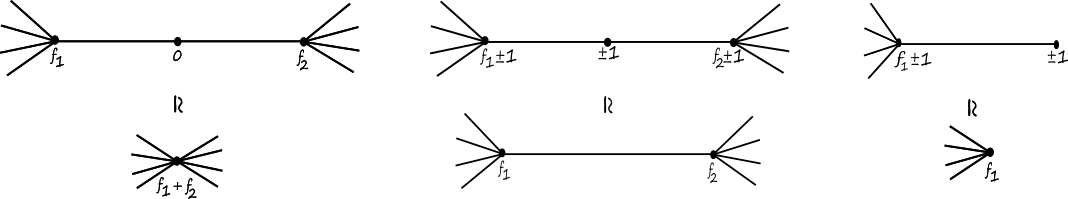}
	\caption{Kirby-Neumann moves that relate plumbing graphs which result in homeomorphic 3-manifolds.}
	\label{fig:moves}
\end{figure}

Let us confine to the   three-manifold $M[\Gamma]$ obtained from surgery of framed link associated with $L$-vertex graph (an example illustrated in Figure \ref{fig:plumbing-example}). These kind of manifolds are known in the literature as plumbed three-manifolds. The linking matrix $B$ is defined as 
\begin{equation}
	B_{v_1,v_2}=\left\{
	\begin{array}{ll}
		1,& v_1,v_2\text{ connected}, \\
		f_v, & v_1=v_2=v, \\
		0, & \text{otherwise}.
	\end{array}
	\right.\qquad v_i \in \text{Vertices of }\Gamma \;\cong\;\{1,\ldots,L\}.
\end{equation}
The algebraic expression for the WRT invariant $\tau_k^{\mathcal G}[M(\Gamma);\q]$ is
\begin{equation}
	\tau_k^{\mathcal G}[M(\Gamma);\q]=\frac{F^{\mathcal G}[\CL(\Gamma);\q]}{F^{\mathcal G} [\CL(+1\bullet);\q]^{b_+}F^{\mathcal G} [\CL(-1\bullet);\q]^{b_-}}
	\label{WRT}
\end{equation} 
where $b_{\pm}$ are the number of positive and negative eigenvalues of a linking matrix $B$ respectively and $F^{\mathcal{G}}[\mathcal{L}(\Gamma);\q]$ is defined as
\begin{equation}
	F^{\mathcal G}[\CL(\Gamma);\q] = \sum_{R_1,R_2,\ldots R_L}\left( \prod_{i=1}^L V_{R_i}^{\mathcal G} [\bigcirc;\q]\right)  V_{R_1,R_2\ldots R_L}^{\mathcal G} [\mathcal L(\Gamma);\q]~,
\end{equation}
where the summation indicates all the allowed integrable representations of affine $\mathfrak{g}_k$ Lie algebra. By construction, any two homeomorphic manifolds must share the same three-manifold invariant. There is a prescribed set of moves called Kirby moves on links which gives the same three-manifold. For framed links depicted as plumbing graphs, these moves are known as Kirby-Neumann moves as shown in Figure~\ref{fig:moves}.
Hence,  the three-manifold invariant must obey
\begin{equation}
	\tau_k^{\mathcal G}[M(\Gamma);\q]=
	\tau_k^{\mathcal G}[M(\Gamma');\q]~,
\end{equation}
where the plumbing graphs $\Gamma, \Gamma'$ are related by the Kirby-Neumann moves.

Towards the end of 20th century, attempts to give a topological interpretation for the integer coefficients  in the Laurent series expression for Jones polynomial (HOMFLY-PT)  as well as the corresponding colored invariants  for any knot  $\mathcal K$
\begin{equation}
	J_{n}[\mathcal K;\q]  = \sum_s  a_s \q^s~, ~ \{ a_s \} \in \mathbb Z \label {laurent}
\end{equation}
has resulted in developments on homology theories as well as physics explanation.
We will discuss these  `homological invariants'  and their appearance in string/M-theory  in the 
following section.
\subsection{Knot, Link and Three-manifold Homologies}
We will first review the developments on homological invariants of knots accounting for these integers $a_s$ (\ref{laurent}) as  dimension of the vector space $\mathcal H_{\mathcal K}$ of a homological  theory. Then, we will present the topological string/M-theory approach where these integers count number of BPS states. 
\subsubsection{Homological Invariants of Knots}
The pioneering work of Khovanov\cite{khovanov2000categorification} on bi-graded homology theory led to categorification of  the Jones polynomial. This was extended to colored $\mathfrak{sl2}$  knot homology $\mathcal H^{\mathfrak {sl}_2;n}_{i,j}$ \cite{webster2017knot,cooper2012categorification,frenkel2012categorifying} leading to new homological invariants $\mathcal P_n^{\mathfrak{sl}_2}[\mathcal K,q,t]$ which categorifies the colored  Jones polynomial: 
\begin{equation}
	\mathcal P_n^{\mathfrak{sl}_2}[\mathcal K,\q,t]=\sum_{i,j} t^j   \q^i {\rm dim} \mathcal {H}^{\mathfrak {sl}_2;n}_{i,j}~. \label{sl2homo}
\end{equation}
The subscripts $i$ and $j$ on the colored $\mathfrak {sl}_2$ homology $\mathcal{H}^{\mathfrak {sl}_2;n}_{i,j}$ are called the polynomial  grading and the homological grading respectively. In fact, the $\q$-graded Euler characteristic of the colored $\mathfrak {sl}_2$ knot homology gives the colored Jones invariant:
\begin{equation} 
	J_{n}[\mathcal K;\q] = \sum_{i,j} (-1)^j   \q^i {\rm dim} \mathcal H^{\mathfrak {sl}_2;n}_{i,j}~,
\end{equation} 
explaining the reasons behind the integers $a_s$(\ref {laurent}).
Khovanov and Rozansky \cite{khovanov2004matrix} constructed  $\mathfrak {sl}_N$ homology using matrix factorizations. This  led to the categorification of colored HOMFLY-PT polynomials of knots.
There has been interesting insight on these homological invariants within  topological strings context and  $M$-theory. We will now discuss the essential features from physics approach.
\subsubsection{Topological Strings and M-theory}
The  parallel  developments from topological strings and intersecting branes in $M$-theory \cite{Ooguri:1999bv,Gopakumar:1998jq} interpreted the integers of unnormalised HOMFLY-PT  (\ref{laurent})   as counting of BPS states. Invoking topological string duality in the presence of any knot $\mathcal K$, Ooguri-Vafa conjectured 
\begin{equation} 
	V_{\tiny \yng(1)}^{SU(N)}[\mathcal K;\q,\lambda=\q^N]= {1 \over (\q^{1/2} - \q^{-1/2})} \sum_{Q,s} N_{\tiny \yng(1), Q,s} \lambda^Q \q^s~,
\end{equation}
where the integers $N_{\tiny \yng(1), Q,s}$ count $D4-D2$ bound states in string theory.
Further the relation between the  BPS spectrum and $sl_N$/Khovanov-Rozansky knot homology was conjectured within the topological string context\cite{Gukov:2004hz} :
\begin{equation}
	N_{\tiny\yng(1), Q,s}= \sum_j (-1) ^j  D_{Q,s,j}~,
\end{equation} 
where the integers $D_{Q,s,j}$ are referred to as refined BPS invariants. The extra charge/ homological grading $j$ are explainable by the appearance of extra $U(1)$ symmetry in $M$-theory compactified on Calabi-Yau three-folds $CY_3$.  The topological string duality and the   dualities of  physical string theories compactified on $CY_3$ implies that the vector space of knot homologies are the Hilbert space of BPS states(see review \cite{Nawata:2015wya} and references therein):
$$\mathcal H_{\mathcal K} \equiv \mathcal H_{BPS}~.$$

The impact of knot homology on the categorification of the WRT invariants has been studied in the last six years. We now present a concise summary of the recent developments in this direction.
\subsubsection{Three-Manifold Homology}
\label{sec2.2.3}
As WRT invariants (\ref{WRT}) of three-manifolds  involves  invariants of links, logically we would expect the homology of three-manifold  $\mathcal H^{\mathcal G;M}$  such that
\begin{equation}
	\tau_k^{\mathcal G}[M;\q] 
	\stackrel{\text{?}}{=}
	\sum_{i,j} (-1)^j \q^i {\rm dim} \mathcal H^{\mathcal G; M}_{i,j}~.\label {qhatz}
\end{equation}
However, the WRT invariants known for many three-manifolds are not seen as $\q$-series (\ref{qhatz}). 
We will now review the necessary steps \cite{Gukov:2017kmk} of obtaining a new three-manifold invariant $\hat Z$, as $\q$-series, from  $U(N)$ Chern-Simons partition function for Lens space $M=L(p,1)\equiv S^3 / \mathbb Z_p$. The space of flat connections  $\{a\}$ denoted by $\pi_1[{S^3 \over \mathbb Z_p}]\equiv \mathbb Z_p$. Hence $Z_k^{U(N)}[L[p,1];\q]$  can be decomposed as  a sum of perturbative Chern-Simons $Z_a^{U(N)}[L[p,1];\q]$  around these abelian flat connections $a$ \cite{Gukov:2017kmk}:
\begin{equation}
	Z_k^{U(N)}[L[p,1];\q]= \sum_a \exp [i S_{CS}^{(a)} ]Z_a^{U(N)} [L[p,1];\q]~,
	\label{eqn2.26}
\end{equation}
where $S_{CS}^{(a)}$ is the corresponding classical Chern-Simons action. The following change of basis by $\mathcal S$ matrix of $\mathfrak u(1)^N_p$ affine algebra:
\begin{equation} 
	Z_a^{U(N)}[L(p,1);\q] = \sum_b \mathcal S_{ab} \hat Z_b^{U(N)}[\mathcal L[p,1];q]\Big\vert_{q\rightarrow \q}~,
	\label{eqn2.27} 
\end{equation} 
is required  so that 
\begin{equation} 
	\hat Z_b^{U(N)}[\mathcal L[p,1];q] \in q^{\Delta_b} \mathbb Z[[q]]~,~~ \Delta_b \in \mathbb Q~.
	\label{eqn2.28}
\end{equation}
Physically,  the  $\hat Z_b[\mathcal L[p,1];q]$  is  also the vortex partition function $\hat Z_{T[L[p,1]]}[D^2 \times_q S^1]$ obtained from reducing 
6d $(2,0)$ theory (describing dynamics of $N$-coincident $M5$ branes on $L[p,1] \times D^2 \times_q S^1$) on $L[p,1]$. The effective 3-d $\mathcal N=2$ theory on $D^2 \times_q S^1$ (cigar geometry) is denoted as $T^{U(N)} [L[p,1]]$.  

For other three-manifolds $M$, $\mathcal S$ matrix depends only on $H_1(M,\mathbb{Z})$. 
Further the  Hilbert space of BPS states $\mathcal H^{i,j}_{BPS}$  on the M5 brane system, in the ambient space-time $T^*M  \times TN \times S^1$,  where $i,j$ gradings will keep track of both spins  associated with the rotational symmetry $U(1)_q \times U(1)_R$ on $D^2\subset TN$. The Hilbert space of states for the theory $T^{\mathcal G}[M]$ with boundary condition at $\partial D^2= S^1$  labeled by $a \in ({\rm Tor} H_1(M, \mathbb Z))^N/S_N$ leads to  bi-graded homological invariants of $M$:
\begin{equation}
	{\mathcal H}_a^{U(N)} [M]= {\mathcal H}_{T_{L[p,1]}^{U(N)}} [D^2; a]
	= \bigoplus_{\substack{i \in \mathbb Z+\Delta_a,\\j \in \mathbb Z}}  {\mathcal H}_a^{i,j}~.
\end{equation}
Note that the grading $i$ counts the charge under $U(1)_q$ rotation of $D^2$ and homological grading $j$  is the R-charge of the $U(1)_R$ R-symmetry. In the following section, we will review the necessary steps of obtaining $\hat Z$ invariants for $SU(2)$ group. This will provide clarity of notations to investigate $\hat Z$ for  $SO(3)$ and $OSp(1|2)$ group.

\section{Review of $SU(2)$ $\hat{Z}$ invariant}
\label{reviewsu2}
As discussed in subsection \ref{sec2.2.3}\cite{Gukov:2016gkn},  the expression for Lens space partition function using eqns.(\ref{eqn2.26}-\ref{eqn2.28})
\begin{eqnarray}
	Z_k^{U(N)}[L(p,1),\q]&=&\sum_{a,b\in \mathbb{Z}_p}\mathcal{S}_{ab} \exp[iS_{CS}^{(a)}]\hat{Z}_b^{U(N)}[\mathcal{L}(p,1);q] \Big\vert_{q \rightarrow \q}~,\\
	{\rm where}~	\hat Z_b[\mathcal L[p,1];q] &\in& q^{\Delta_a} \mathbb Z[[q]]~,~~ \Delta_a \in \mathbb Q~.
\end{eqnarray}
led to the following  conjecture \cite{Gukov:2017kmk,Gukov:2019mnk} for any closed oriented  three manifold $M$ known as GPPV conjecture:
\begin{eqnarray}
	Z^{SU(2)}_k[M;\q]&=&(i\sqrt{2(k+2}))^{b_1(M)-1}\sum_{a,b\;\in \;\atop \text{Spin}^c(M)/\Z_2}\exp[2\pi i(k+2) \lk(a,a)]
	\,\times\nonumber\\
	~&~&~~~~~~~|\mathcal{W}_b|^{-1}\mathcal{S}_{ab}
	\hat{Z}^{SU(2)}_b[M;q]|_{q\rightarrow \q=\exp({\frac{2\pi i }{k+2}})}
	\label{WRT-decomposition3.2}
\end{eqnarray}
where
\begin{equation}
	\hat{Z}^{SU(2)}_b[M;q] \in \, 2^{-c} q^{\Delta_b} \Z[[q]]\qquad \Delta_b\in \Q,\qquad c\in\Z_+
	\label{Block-qSeries}
\end{equation}
is convergent for $|q|<1$ and
\begin{equation}
	\mathcal{S}_{ab}=\frac{e^{2\pi i\lk(a,b)}+e^{-2\pi i\lk(a,b)}}{|\CW_a|\sqrt{|H_1(M,\Z)|}}.
	\label{Sab}
\end{equation}
Here $\mathcal{W}_a$ is the stabilizer subgroup defined as 
\begin{equation}
	\CW_a \; \equiv \; \text{Stab}_{\Z_2}(a) \; = \; \left\{
	\begin{array}{cl}
		\Z_2, & a=-a \,, \\
		1, & \text{otherwise\,.}
	\end{array}
	\right.
\end{equation}
and $\ell k$ denotes the  linking pairing on $H_1(M,\mathbb{Z})$:

\begin{equation}
	\begin{array}{cccc}
		\lk: & H_1(M,\mathbb{Z})\otimes H_1(M,\mathbb{Z}) & \longrightarrow & \Q/\Z \\
		& [a]\otimes [b] & \longmapsto & {\#(a\cap \hat b)}/{n} \\
	\end{array}
\end{equation}
where $\hat b$ is a two-chain complex such that $\partial \hat b = nb$ with $n\in \Z$. Such a 
$\hat b$ and $n$ exists because $[b] \in H_1(M,\mathbb{Z})$.  The number $\#(a\cap \hat b)$ counts the  intersection points with signs determined by the orientation. The set of orbits is the set of $\text{Spin}^c$ structures on $M$, with the action of $\Z_2$ by conjugation.

Although, the relation(\ref{WRT-decomposition3.2}) is true for any closed oriented three-manifold $M$, the explicit $q$ series expression for $\hat{Z}$ is waiting to be discovered for a general three-manifold. 

In the following subsection, we will review the $\hat{Z}^{SU(2)}$  for the  plumbed manifolds. We begin with the WRT invariant for a plumbing graph, of the type shown in Figure. \ref{fig:plumbing-example}, discussed in section.(\ref{sec2.1.2}). Then analytically continue $\q \rightarrow q$ to get the $\hat{Z}^{SU(2)}$-invariant . We will see that  the analytic continuation  procedure is doable only for negative definite plumbed manifolds({\it i.e.,} the signature of linking matrix $B$, $\sigma= b_+ - b_-= -L$)\footnote{In principle, this procedure is also doable when $B$ is negative on a certain subspace of $\mathbb{Z}^L$.}. Moreover, as explained in\cite{Gukov:2019mnk}, the $\text{Spin}^c$-structure in case of plumbed 3-manifold with $b_1(M)=0$, is given by $H_1(M,\mathbb{Z})\cong \text{Coker} B=\mathbb{Z}^L/B\mathbb{Z}^L$.
\subsection{$\hat{Z}_b^{SU(2)(q)}$}
The WRT invariant $\tau_k^{SU(2)}[M(\Gamma);\q]$,\footnote{normalized such that $\tau_k^{\mathcal{G}}[S^3;\q]=1$ and $k$ is the bare level for $SU(2)$ Chern-Simons theory} for plumbed three-manifold $M(\Gamma)$(\ref{WRT}), obtained from surgery of framed link $\CL(\Gamma)$ in $S^3$, is
\begin{eqnarray}
	\tau_k^{SU(2)}[M(\Gamma);\q]&=&\frac{F^{SU(2)}[\CL(\Gamma);\q]}{F^{SU(2)} [\CL(+1\bullet);\q]^{b_+}F^{SU(2)}[\CL(-1\bullet);\q]^{b_-}}
	\nonumber\\
	{\rm where}~F^{SU(2)}[\CL(\Gamma);\q]&=& \sum_{{n}\in \{1,\ldots,k+1\}^{L}} J[\CL(\Gamma)]_{n_1,\ldots,n_L}
	\prod_{v=1}^L \frac{\q^{n_v/2}-\q^{-n_v/2}}{\q^{1/2}-\q^{-1/2}}.
\end{eqnarray}
Note $b_{\pm}$ are the number of positive and negative eigenvalues of a linking matrix $B$ respectively  and the colored Jones polynomial of link $\mathcal{L}(\Gamma)$ (\ref{conn1}) in variable $\q = \exp({2 i \pi /(k+2)})$ is
\begin{eqnarray}
	J[\CL(\Gamma)]_{n_1,\ldots,n_L}&=&\frac{2i}{\q^{1/2}-\q^{-1/2}}\prod_{v\;\in\; \text{Vertices}\;\cong\;\{1,\ldots,L\}}
	\q^{\frac{f_v(n_v^2-1)}{4}}
	\,
	\times \\
	~&~&\left(\frac{2i}{\q^{n_v/2}-\q^{-n_v/2}}\right)^{\text{deg}(v)-1}\prod_{(v_1,v_2)\;\in\;\text{Edges}}
	\frac{\q^{n_{v_1}n_{v_2}/2}-\q^{-n_{v_1}n_{v_2}/2}}{2i}.\nonumber
\end{eqnarray}
Using the following Gauss sum reciprocity formula 
\begin{multline}
	\sum_{n\;\in\;\Z^L/2k\Z^L}
	\exp\left(\frac{\pi i}{2k}(n,Bn)+\frac{\pi i}{k}(\ell,n)\right)
	=\\
	\frac{e^{\frac{\pi i\sigma}{4}}\,(2k)^{L/2}}{|\det B|^{1/2}}
	\sum_{a\;\in\; \Z^L/B\Z^L}
	\exp\left(-2\pi i k\left(a+\frac{\ell}{2k},B^{-1}\left(a+\frac{\ell}{2k}\right)\right)\right)
	\label{reciprocity11}
\end{multline}
where $\ell \in \Z^L$, $(\cdot,\cdot)$ is the standard pairing on $\Z^L$ and $\sigma=b_+-b_-$ is the signature of the linking matrix $B$, we can sum 
\begin{equation}
	F^{SU(2)}[\CL(\pm 1\bullet); \q]=\sum_{n=1}^{k+1}\q^{\pm\frac{n^2-1}{4}}\,
	\left(\frac{\q^{n/2}-\q^{-n/2}}{\q^{1/2}-\q^{-1/2}}\right)^2=
	\frac{[2(k+2)]^{1/2}\,e^{\pm\frac{\pi i}{4}}\,\q^{\mp\frac{3}{4}}}{\q^{1/2}-\q^{-1/2}},
\end{equation}
for the unknot with framing $\pm 1$.  Incorporating the above equation and the fact that $L-|\text{Edges}|=1$ for  the framed link $\CL(\Gamma)$, the WRT invariant simplifies to 
\begin{multline}
	\tau_k^{SU(2)}[M(\Gamma);\q]=\frac{e^{-\frac{\pi i\sigma}{4}}\,\q^{\frac{3\sigma}{4}}}{2\,(2(k+2))^{L/2}\,(\q^{1/2}-\q^{-1/2})}
	\times\\
	{\sum_{{n}\in \Z^L/2(k+2)\Z^L}}'\prod_{v\;\in\; \text{Vertices}}
	\q^{\frac{f_v(n_v^2-1)}{4}}
	\,\left(\frac{1}{\q^{n_v/2}-\q^{-n_v/2}}\right)^{\text{deg}(v)-2}
	\times \\
	\prod_{(v',v'')\;\in\;\text{Edges}}
	\frac{\q^{n_{v'}n_{v''}/2}-\q^{-n_{v'}n_{v''}/2}}{2}
	\label{WRT-Gamma-01}
\end{multline}
where we used invariance of the summand under $n_v \rightarrow -n_v$. The prime $'$ in the sum means that the singular values $n_v=0,\,k+2$ are omitted. Let us focus on the following factor for general plumbed graph:
\begin{eqnarray}
	\prod_{(v',v'')\;\in\;\text{Edges}}
	(\q^{n_{v'}n_{v''}/2}-\q^{-n_{v'}n_{v''}/2})
	&=&\sum_{{p}\in\{\pm 1\}^\text{Edges}}
	\prod_{(v',v'')\;\in\;\text{Edges}}p_{(v',v'')} \nonumber\\
	~&~&\q^{p_{(v',v'')}n_{v'}n_{v''}/2}.
\end{eqnarray}
Note that, under $n_v \rightarrow -n_v$ on any vertex  $v$ of degree ${\rm deg}(v)$,
the factor  with a given configuration of signs associated to edges ({\it i.e.,} $p\in\{\pm 1\}^\text{Edges}$) will transform into a term with a different configuration times $(-1)^{\text{deg}(v)}$.  For the class of graphs $\Gamma$ (like Figure. \ref{fig:plumbing-example}),  the  sequence of such transforms can be finally  brought to the configuration with all signs $+1$.  Hence, the WRT invariant (\ref{WRT-Gamma-01}) for these plumbed three-manifolds can be reduced to this form:
\begin{multline}
	\tau_k^{SU(2)}[M(\Gamma)]=\frac{e^{-\frac{\pi i\sigma}{4}}\,\q^{\frac{3\sigma-\sum_v f_{v}}{4}}}{2\,(2(k+2))^{L/2}\,(\q^{1/2}-\q^{-1/2})}
	\times\\
	{\sum_{{n}\in \Z^L/2(k+2)\Z^L}}'
	\;\; \q^{\frac{(n,Bn)}{4}}
	\prod_{v\;\in\; \text{Vertices}}
	\,\left(\frac{1}{\q^{n_v/2}-\q^{-n_v/2}}\right)^{\text{deg}(v)-2}.
	\label{WRT-Gamma-11}
\end{multline}

In the above expression, the points $0$ and $k+2$ are excluded in the summation but in the reciprocity formula (\ref{reciprocity11}) no point is excluded. So, to apply the reciprocity formula  we have to first regularize the sum. This is achieved by  introducing the following regularising parameters:
\begin{eqnarray}
	\Delta_v\in\Z_+:\; \Delta_v&=&\text{deg}(v) -1\mod 2,\qquad \forall v\;\in\; \text{Vertices},
	\label{Delta-def}\\
	\omega\in \C:&\;\;&0<|\omega|<1.\nonumber
\end{eqnarray}
so that the sum in eqn.(\ref{WRT-Gamma-11}) is rewritable as  $\omega \rightarrow  1$:
\begin{multline}
	{\sum_{{n}\in \Z^L/2(k+2)\Z^L}}'
	\;\; \q^{\frac{(n,Bn)}{4}}
	\prod_{v\;\in\; \text{Vertices}}
	\,\left(\frac{1}{\q^{n_v/2}-\q^{-n_v/2}}\right)^{\text{deg}(v)-2}=\\
	\lim_{\omega\rightarrow 1} \frac{1}{2^L}\sum_{{n}\in \Z^L/2(k+2)\Z^L}
	\q^{\frac{(n,Bn)}{4}}
	F_\omega(x_1,\ldots,x_L)|_{x_v=\q^{n_v/2}},
	\label{WRT-omega-limit}
\end{multline}
where
\begin{eqnarray}
	F_\omega(x_1,\ldots,x_L)&=&
	\prod_{v\;\in\; \text{Vertices}}\left({x_v-1/x_v}\right)^{\Delta_v}
	\times\,\\
	\label{F-omega}
	~&~&\left\{
	\left(\frac{1}{x_v-\omega/x_v}\right)^{\text{deg}(v)-2+\Delta_v}
	+ \left(\frac{1}{\omega x_v-1/x_v}\right)^{\text{deg}(v)-2+\Delta_v}
	\right\}\nonumber
	\end{eqnarray}
Note that, we can perform a binomial expansion taking $(\omega/x_v^2)$ small in the first term and 
$(\omega x_v^2)$ small in the second term to rewrite $F_\omega(x_1,\ldots,x_L)$ as a formal power series:
\begin{equation}
	F_\omega(x_1,\ldots,x_L)=\sum_{\ell\in 2\Z^L +\delta} F_\omega^\ell\prod_v x_v^{
		\ell_v}
	\qquad \in \Z[\omega][[x_1^{\pm1},\ldots,x_1^{\pm L}]]~,
\end{equation}
where $\delta\in \Z^L/2\Z^L,~ \delta_v \equiv \text{deg}(v)\mod 2$ and 
 \begin{equation}
	F_\omega^\ell=\sum_{m:\,\ell\in \CI_m}N_{m,\ell}\,\omega^m \;\in \Z[\omega]
\end{equation}
with $\CI_m$ being a finite set of elements from $\Z^L$.
By definition, ${\rm lim}_{\omega \rightarrow 1} F_{\omega}^\ell$ is not dependent on $\Delta\in \Z^L$  (\ref{Delta-def}). 
However this $\omega \rightarrow 1$ limit in eqn. (\ref{F-omega}) will restrict the binomial expansion range of the first term to be $x \rightarrow \infty$ and that of the second term to $x \rightarrow 0$:
\begin{eqnarray}
	F_{\omega \rightarrow 1} (x_1,\ldots,x_L)=\sum_{\ell\in 2\Z^L +\delta} F_{\omega \rightarrow 1}^\ell\prod_v x_v^{\ell_v}&=& \\
	\label{F1lsu2}
	~~{\rm lim}_{\omega \rightarrow 1}\prod_{v\,\in\,\text{Vertices}}\left\{
	{\scriptsize \begin{array}{c} \text{Expansion} \\ \text{as} x\rightarrow \infty \end{array} }
	\frac{1}{(x_v-\omega/x_v)^{\text{deg}\,v-2}}\right.
	&+& \left . {\scriptsize \begin{array}{c} \text{Expansion} \\ \text{as }  x\rightarrow 0 \end{array} }
	\frac{1}{(\omega x_v-1/x_v)^{\text{deg}\,v-2}}\right\}.\nonumber	
\end{eqnarray}
Now let us assume that the quadratic form $B:\Z^L\times \Z^L\rightarrow \Z$ is negative definite {\it i.e., } $\sigma=-L$. Then we can define the following series in $q$ which is convergent for $|q|<1$:
\begin{equation}
	\hat Z_b^{SU(2)}[M(\Gamma);q]\stackrel{\text{Def}}{=\joinrel=}
	2^{-L} q^{-\frac{3L+\sum_v f_{v}}{4}}
	\sum_{\ell \in 2B\Z^L+b}F^\ell_{\omega \rightarrow 1}\,q^{-\frac{(\ell,B^{-1}\ell)}{4}}
	\;\in\; 2^{-c}q^{\Delta_b}\Z[[q]]
	\label{Z-hat-def}
\end{equation}
where $c\in \Z_+, c\leq L$ and 
\begin{eqnarray}
	b&\in& (2\Z^L+\delta)/2B\Z^L\,/\Z_2 \cong (2\text{Coker}\,B+\delta)\,/\Z_2
	\stackrel{\text{Set}}{\cong} H_1(M_3,\Z)\,/\Z_2,\\
	\Delta_b&=&-\frac{3L+\sum_v f_{v}}{4}-\max_{\ell \in 2M\Z^L+b}\frac{(\ell,B^{-1}\ell)}{4}\,\in \Q
\end{eqnarray}
where $\Z_2$ action takes $b\rightarrow -b$ and is the symmetry of (\ref{Z-hat-def}).

Using relation (\ref{WRT-omega-limit}) and applying Gauss reciprocity formula (\ref{reciprocity11}) we arrive at the following expression for the WRT invariant:
\begin{multline}
	\tau_{k}^{SU(2)}[M(\Gamma);\q]=\\\\ ~~~~~\frac{e^{-\frac{\pi iL}{4}}\,\q^{-\frac{3L+\sum_v f_{v}}{4}}}{2\,(2(k+2))^{L/2}\,(\q^{1/2}-\q^{-1/2})}\,
	\lim_{\omega\rightarrow 1}\sum_{{n}\in \Z^L/2(k+2)\Z^L}
	\q^{\frac{(n,Bn)}{4}}
	F_\omega(x_1,\ldots,x_L)|_{x_v=\q^{n_v/2}}
	\label{WRT-Gamma-2}=\\\\
	~~\frac{2^{-L} \q^{-\frac{3L+\sum_v f_{v}}{4}}}{2\,(\q^{1/2}-\q^{-1/2})\,|\det B|^{1/2}}
	\sum_{\scriptsize\begin{array}{c}a\in \mathrm{Coker}\,B \\ b \in 2\mathrm{Coker}\, B+\delta \end{array}}
	e^{-2\pi i(a,B^{-1}b)} e^{-2\pi i(k+2)(a,B^{-1}a)}\times\\
	\lim_{\omega\rightarrow 1}
	\sum_{\ell \in 2B\Z^L+b}F^\ell_\omega\,\q^{-\frac{(\ell,B^{-1}\ell)}{4}}.
\end{multline}
Assuming that the limit $\lim_{q\rightarrow \q}\hat{Z}_b^{SU(2)}(q)$ exists, where  $q$ approaches $(k+2)$-th primitive root of unity from inside of the unit disc $|q|<1$, we expect
\begin{equation}
	\lim_{\omega\rightarrow 1}
	\sum_{\ell \in 2B\Z^L+b}F^\ell_\omega\,\q^{-\frac{(\ell,B^{-1}\ell)}{4}}
	=\lim_{q\rightarrow \q}
	\sum_{\ell \in 2B\Z^L+b}F^\ell_{\omega \rightarrow 1}\,q^{-\frac{(\ell,B^{-1}\ell)}{4}}.
	\label{limit-exchange}
\end{equation}
Thus we obtain GPPV conjecture form:
\begin{multline}
	\tau_{k}^{SU(2)}[M(\Gamma),\q]=\frac{1}{2\,(\q^{1/2}-\q^{-1/2})\,|\det B|^{1/2}}
	\,\times\\
	\sum_{a\in \mathrm{Coker}\,B}e^{-2\pi i(k+2)(a,B^{-1}a)}
	\sum_{b \in 2\mathrm{Coker}\, B+\delta} e^{-2\pi i(a,B^{-1}b)}\lim_{q\rightarrow \q} \hat{Z}_b^{SU(2)}[M(\Gamma);q].
	\label{WRT-prop-1}
\end{multline}
There is also an equivalent  contour integral form for  the homological blocks(\ref{Z-hat-def}):\begin{equation}
	\hat{Z}_b^{SU(2)}[M(\Gamma);q]=q^{-\frac{3L+\sum_v f_{v}}{4}}\cdot\text{v.p.}\int\limits_{|z_v|=1}
	\prod_{v\;\in\; \text{Vertices}}
	\frac{dz_v}{2\pi iz_v}\,
	\left({z_v-1/z_v}\right)^{2-\text{deg}(v)}\cdot\Theta^{-B}_b(z),
\end{equation}
where $\Theta^{-B}_b(x)$ is the theta function of the lattice corresponding to minus the linking form $B$:
\begin{equation}
	\Theta^{-B}_b(x)=\sum_{\ell \in 2B\Z^L+b}q^{-\frac{(\ell,B^{-1}\ell)}{4}}
	\prod_{i=1}^Lx_i^{\ell_i},
\end{equation}
and ``v.p.'' refers to principle value integral (\ie~take half-sum of contours $|z_v|=1\pm \epsilon$). This prescription corresponds to the regularization by $\omega$ made in eqn.(\ref{F-omega}).

Thus we can obtain explicit  $SU(2)$ $q$-series for any negative definite plumbed three-manifolds. For completeness, we  present the $q$-series for some examples. 
\subsection{Examples}
$\bullet$ Poincare homology sphere is a well-studied three-manifold corresponding to the graph:
\begin{equation}
	\includegraphics{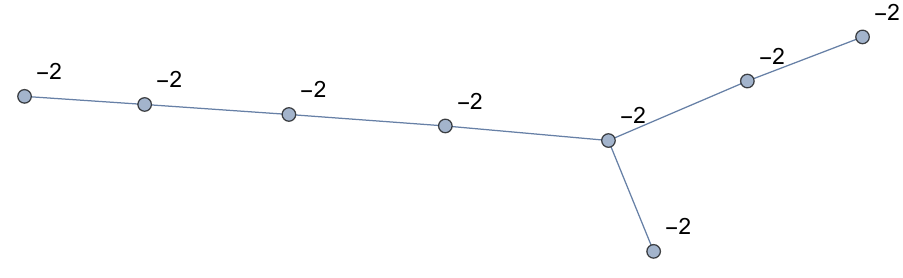}
	\label{eqn3.31new}
\end{equation}
As $H_1(M,\Z)=0$, we obtain only single homological block $\hat Z_{b_1}$. Solving eqns.(\ref{F1lsu2},\ref{Z-hat-def}), we get
\begin{equation}
	\hat{Z}_{b_1}^{SU(2)}=q^{-3/2}(1-q-q^3-q^7+q^8+q^{14}+q^{20}+q^{29}-q^{31}-q^{42}+\cdots).
\end{equation}
$\bullet$ The next familiar example with $H_1(M,\Z)=0$ is 
Brieskorn homology sphere. A particular example of this class is $\Sigma(2,3,7)$ with the following equivalent graphs:
\begin{equation}
	{\,\raisebox{-2.25cm}{\includegraphics[width=4.5cm]{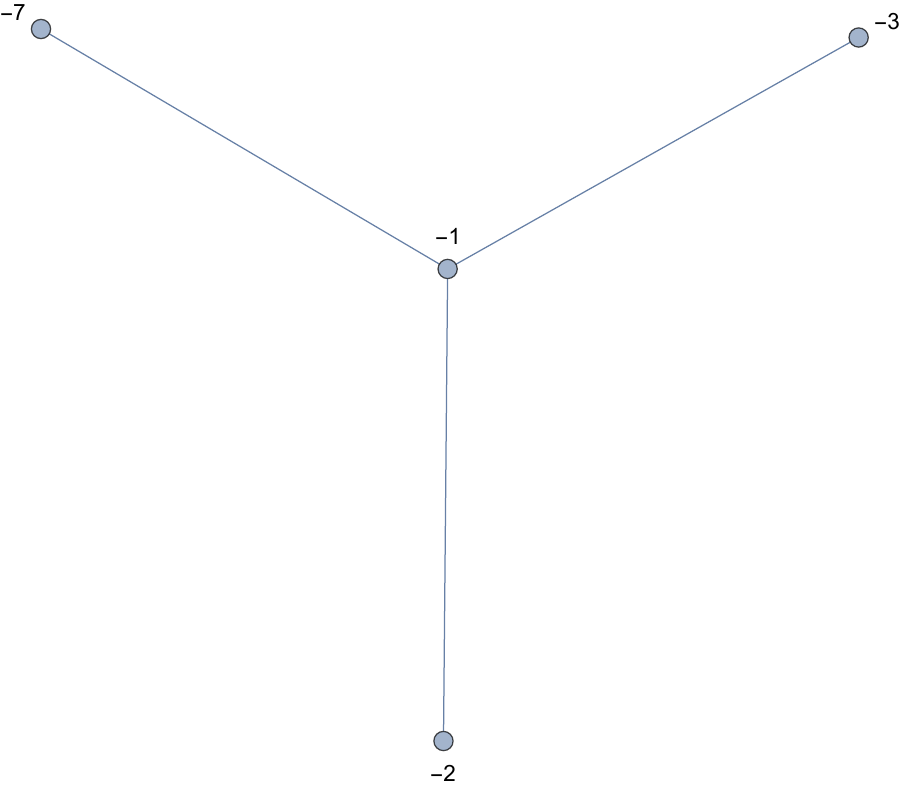}}\,}
	\stackrel{\text{Kirby}}{\sim}
	{\,\raisebox{-2.1cm}{\includegraphics[width=3.75cm]{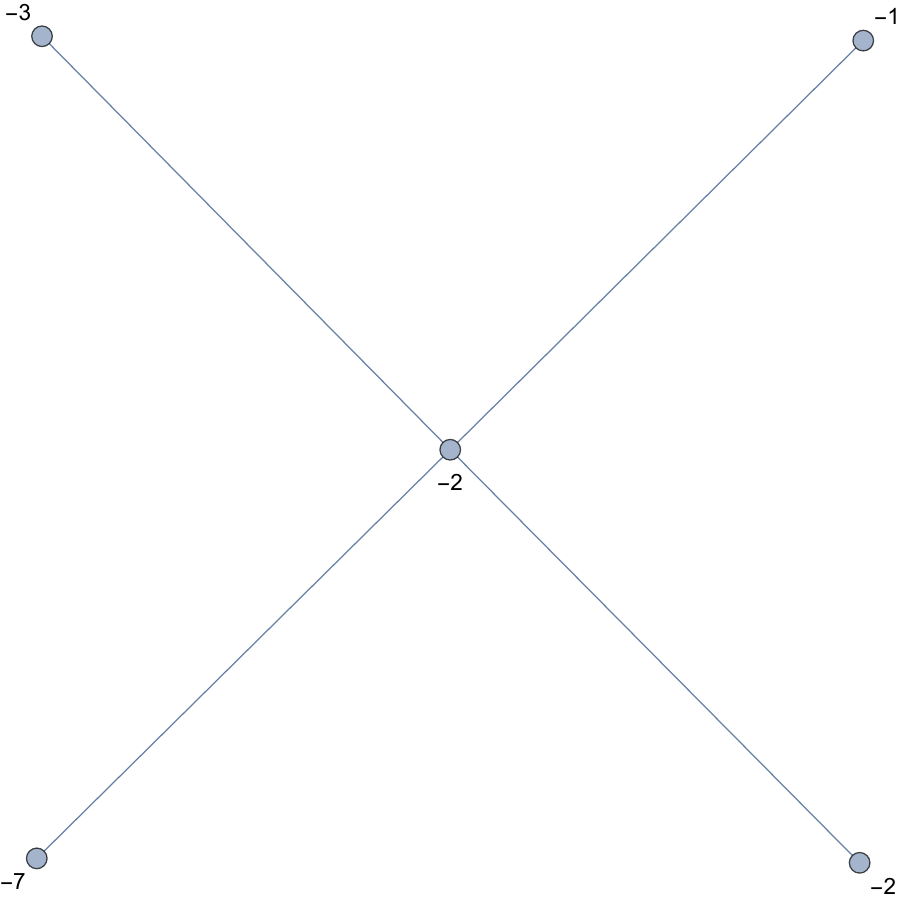}}\,}
	\label{eqn3.33new}
\end{equation}
The homological block turns out to be
\begin{equation}
	\hat{Z}_{b_1}^{SU(2)}=q^{1/2}(1 - q - q^5 + q^{10} - q^{11} + q^{18} + q^{30} - q^{41} + q^{43} - q^{56} - 
	q^{76}\cdots).
\end{equation}
$\bullet$ For a three-manifold with non-trivial $H_1(M,\Z)=\Z_{3}$ as drawn below, 
\begin{equation}
	{\,\raisebox{-2.0cm}{\includegraphics[width=8.0cm]{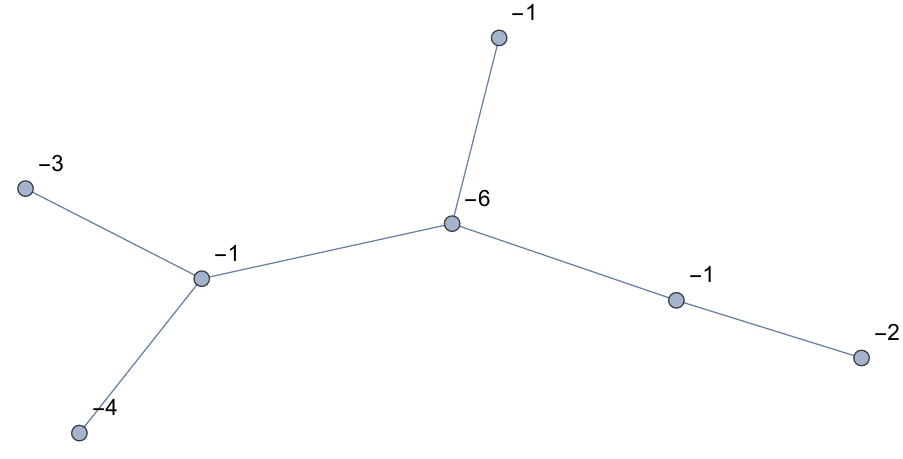}}\,}
	\stackrel{\text{Kirby}}{\sim}
	{\,\raisebox{-2.0cm}{\includegraphics[width=4.0cm]{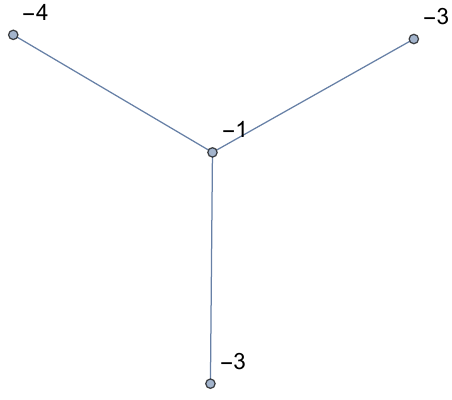}}\,}
	\label{eqn3.35new}
\end{equation}
the three homological blocks are
\begin{equation}
	\hat{Z}^{SU(2)}=
	\left(
	\begin{array}{c}
		1-q+q^6-q^{11}+q^{13}-q^{20}+q^{35}+O\left(q^{41}\right) \\
		q^{5/3}\left(-1+q^3-q^{21}+q^{30}+O\left(q^{41}\right)\right) \\
		q^{5/3}\left(-1+q^3-q^{21}+q^{30}+O\left(q^{41}\right)\right) \\
	\end{array}
	\right),
\end{equation}
where two of them are equal.

Our focus is to obtain explicit $q$-series for $SO(3)$ and $OSp(1|2)$ groups.
Using the relation between $SU(2)$ and $SO(3)$, $SU(2)$ and $OSp(1\vert 2)$ link invariants(\ref{sec2.1.1}),  we will investigate the necessary steps starting from the WRT invariant for $SO(3)$ and $OSp(1|2)$ eventually leading to the  $\hat Z$-invariant. This will be the theme of the following section.

\section{$\hat{Z}$ for $SO(3)$ and $OSp(1\vert2)$}
\label{zhatnew}
Our aim is to  derive the $\hat{Z}$-invariant for $SO(3)$ and $OSp(1\vert 2)$ groups. 
We will first look at  the WRT invariants $\tau_K^{SO(3)}[M(\Gamma); Q]$ for plumbed three-manifolds written in terms of colored Jones invariants of framed links $\CL[\Gamma]$ in the following subsection and then discuss  $OSp(1|2)$ $\hat Z$ in the subsequent section.

\subsection{$SO(3)$ WRT invariant and $\hat{Z}^{SO(3)}$ invariant}
Recall that the framed link invariants are written in  variable $\q$ which is dependent on Chern-Simons coupling and the rank of the gauge group $\mathcal G$. For $SO(3)$ Chern-Simons with coupling $K \in 2 \mathbb Z$, the variable $Q= e^{\frac{2\pi i}{K+1}}$. Hence $F^{SO(3)}[\mathcal{L}(\Gamma);Q]$ in WRT $\tau_K^{SO(3)}[M(\Gamma); Q]$ is

\begin{multline}
	F^{SO(3)}[\mathcal{L}(\Gamma);Q]=\\ \sum_{n_1,n_2,\dots,n_L\in \{0,1,\dots,K\}}V_{n_1,n_2,\dots,n_L}^{SO(3)}(\mathcal{L}(\Gamma);Q)\prod_{v=1}^{L}V_{n_1,n_2,\dots,n_L}^{SO(3)}(\bigcirc;Q)~=\\\\
	\sum_{n_1,n_2,\ldots,n_L\in\{1,3,\dots,2K+1\}}J_{n_1,n_2,\dots,n_L}^{SU(2)}\left(\mathcal{L}(\Gamma); \q=e^{\frac{2\pi i}{2K+2}}\right)\prod_{v=1}^{L}\frac{\q^{n_v/2}-\q^{-n_v/2}}{\q^{1/2}-\q^{-1/2}}\bigg\vert_{\q^2 \rightarrow Q}~,
	\label{eqn4.2}
\end{multline}
where  we have used the relation (\ref{so3}) to write $SO(3)$ link invariants in terms of the colored Jones invariants. Notice that the summation is over only odd integers and hence WRT invariant for $SO(3)$ is different from the WRT for $SU(2)$ group. Further,
the  highest integrable representation in the summation indicates that the Chern-Simons coupling for  $SU(2)$ group is $2K+2$. After performing the summation, we can convert 
the $\q= Q^{1/2}$(\ref{so3}) to obtain $SO(3)$ WRT invariant. We need to modify the Gauss sum reciprocity formula to incorporate the summation over odd integers in $F^{SO(3)}[\mathcal{L}(\Gamma);Q]$.

Using the following  Gauss sum reciprocity formula 
\begin{multline}
	\sum_{n\;\in\;\Z^L/k\Z^L}
	\exp\left(\frac{2\pi i}{k}(n,Bn)+\frac{2\pi i}{k}(\ell,n)\right)
	=\\
	\frac{e^{\frac{\pi i\sigma}{4}}\,(k/2)^{L/2}}{|\det B|^{1/2}}
	\sum_{a\;\in\; \Z^L/2B\Z^L}
	\exp\left(\frac{-\pi ik}{2}\left(a+\frac{\ell}{k},B^{-1}\left(a+\frac{\ell}{k}\right)\right)\right)~,
	\label{reciprocity1}
\end{multline}
for $k=2K+2$,  we can obtain the summation over odd integers by replacing $n \longrightarrow \frac{n-1}{2}$ :
\begin{multline}
	\sum_{n_1,n_2,\dots,n_L\;\in\;\{1,3,\dots,4K+3\}}
	\q^{\frac{(n,Bn)}{4}+\frac{(n,d)}{2}} = \frac{e^{\frac{\pi i\sigma}{4}}\,(K+1)^{L/2}}{|\det B|^{1/2}}
	\q^{-\frac{(d,B^{-1}d)}{4}}\times\\\\ \sum_{a\;\in\; \Z^L/2B\Z^L}
	\exp\left[-\pi i(K+1)(a,B^{-1}a)\right]\exp\left[-\pi i(a,B^{-1}(d+BI))\right],
	\label{reciprocity2}
\end{multline}
where $d = \ell-B I$ with $I$ denoting $L$ component  vector with entry $1$ on all the components. That is, the transpose of the vector $I$ is
\begin{equation}
	I^T=[1,1,\ldots,1]~.\label{eqn4.4ii}
\end{equation}

For unknot with framing $\pm 1$, the $F^{SO(3)}[\CL(- 1\bullet); Q=\q^2]$ involving summation over odd integers simplifies to 
\begin{equation}
	F^{SO(3)}[\CL(\pm\bullet); Q=\q^2]=\frac{\sqrt{K+1}\;e^{\pm\pi i/4}\;\q^{\mp 3/4}}{\q^{1/2}-\q^{-1/2}}\underbrace{(1+e^{\pi i K})}_2~,
\end{equation}
as the coupling $K \in 2 \mathbb Z$ for the $SO(3)$ Chern-Simons theory.
Hence, the WRT invariant takes the following form:
\begin{multline}
	\tau_K^{SO(3)}[M(\Gamma);Q=\q^2]=\frac{e^{-\frac{\pi i\sigma}{4}}\,\q^{\frac{3\sigma}{4}}}{2^L(K+1)^{L/2}\,(\q^{1/2}-\q^{-1/2})}
	\times\\
	{\sum_{{n}\in \{1,3,\ldots,2K+1\}^L}}\prod_{v\;\in\; \text{Vertices}}
	\q^{\frac{f_v(n_v^2-1)}{4}}
	\,\left(\frac{1}{\q^{n_v/2}-\q^{-n_v/2}}\right)^{\text{deg}(v)-2}
	\times \\
	\prod_{(v',v'')\;\in\;\text{Edges}}
	\Big(\q^{n_{v'}n_{v''}/2}-\q^{-n_{v'}n_{v''}/2}\Big).
	\label{WRT-Gamma-4.8}
\end{multline}
In above equation, the terms involving edges of the graph $\Gamma$
\begin{equation*}
	\prod_{(v',v'')\;\in\;\text{Edges}}
	\Big(\q^{n_{v'}n_{v''}/2}-\q^{-n_{v'}n_{v''}/2}\Big)=2^{L-1}		\prod_{(v',v'')\;\in\;\text{Edges}}
	\frac{\Big(\q^{n_{v'}n_{v''}/2}-\q^{-n_{v'}n_{v''}/2}\Big)}{2},
\end{equation*}
can also be rewritten as
\begin{equation*}
	\prod_{(v',v'')\in \text{Edges}} (\q^{n_{v'}n_{v''}/2}-\q^{-n_{v'}n_{v''}/2})=\sum_{p\in\{\pm 1\}^{\text{Edges}}}\prod_{(v',v'')\in \text{Edges}}p_{(v',v'')}\q^{p_{(v',v'')}n_{v'}n_{v''}/2}~.
\end{equation*}
Here again, if we make a change of variable as $n_v\longrightarrow -n_v$ at any vertex, a term in the sum with a given configuration of signs associated to edges (that is $p\in \{\pm 1\}^{\text{Edges}}$) will transform into a term with a different configuration times $(-1)^{\text{deg}(v)}$. However, for these plumbing graphs $\Gamma$, the signs of such configuration can be brought to the configuration with all signs +1. Incorporating this fact, the  WRT invariant(\ref{WRT-Gamma-4.8}) simplifies to 
\begin{multline}
	\tau_{K}^{SO(3)}[M(\Gamma);Q=\q^2]=\frac{e^{-\frac{\pi i\sigma}{4}}\,\q^{\frac{3\sigma-\sum_v f_{v}}{4}}}{2\,(K+1)^{L/2}\,(\q^{1/2}-\q^{-1/2})}
	\times\\
	{\sum_{{n}\in \{1,3,\ldots,2K+1\}^L}}
	\;\; \q^{\frac{(n,Bn)}{4}}
	\prod_{v\;\in\; \text{Vertices}}
	\,\left(\frac{1}{\q^{n_v/2}-\q^{-n_v/2}}\right)^{\text{deg}(v)-2}.
	\label{WRT-Gamma-1}
\end{multline}
\\
Further, we double the range of summation so as to use the reciprocity formula(\ref{reciprocity2})

\begin{multline}
	\tau_{K}^{SO(3)}[M(\Gamma);Q=\q^2]=\frac{e^{-\frac{\pi i\sigma}{4}}\,\q^{\frac{3\sigma-\sum_v f_{v}}{4}}}{4\,(K+1)^{L/2}\,(\q^{1/2}-\q^{-1/2})}
	\times\\
	{\sum_{{n}\in \{1,3,\ldots,4K+3\}^L}}
	\;\; \q^{\frac{(n,Bn)}{4}}
	\prod_{v\;\in\; \text{Vertices}}
	\,\left(\frac{1}{\q^{n_v/2}-\q^{-n_v/2}}\right)^{\text{deg}(v)-2}.
	\label{WRT-Gamma-1so3}
\end{multline}

The steps discussed in the $SU(2)$ context to extract $\hat{Z}$ can be similarly followed  for $SO(3)$. This procedure leads to 
\begin{multline}
	\tau_K^{SO(3)}[M(\Gamma);Q=\q^2]=\frac{1}{2\,(\q^{1/2}-\q^{-1/2})\,|\det B|^{1/2}}
	\,
	\sum_{a\in \mathrm{Coker}\,B}e^{-\pi i(K+1)(a,B^{-1}a)}\\\\
	\sum_{b \in 2\mathrm{Coker}\, B+\delta} e^{-\pi i\big(a,B^{-1}(b{\color{blue}+BI})\big)}\lim_{q\rightarrow \q} \hat{Z}^{SO(3)}_b[M(\Gamma);q]~.
\end{multline}
We observe that the $SO(3)$ WRT invariant is different from the $SU(2)$ invariant due to the factor highlighted in blue color in the summand whereas the $\hat Z^{SO(3)}_b[M(\Gamma);q]$ is exactly same as the $SU(2)$ $q$-series. Even though $SO(3) \equiv SU(2)/\mathbb Z_2$,  it is surprising to see that the factor group $SO(3)$ shares the same $\hat Z$ as that of  the parent group $SU(2)$. The case of $\hat{Z}^{SO(3)}$ was also considered in \cite{Costantino:2021yfd} but they took a different route by considering the refined WRT invariant which is consistent with our result.

In the following subsection, we will extract $\hat Z$ from the WRT invariant $\tau_{\hat K}^{OSp(1|2)}[M(\Gamma); \hat Q]$  for $OSp(1|2)$ supergroup. We will see that the $OSp(1|2)$ $q$-series are related to $\hat Z^{SU(2)}[M(\Gamma);q]$.

\subsection{$OSp(1|2)$ WRT and $\hat{Z}^{OSp(1|2)}$ invariant}

Using the relation between $OSp(1|2)$ and $SU(2)$ link invariants (\ref{osp}), the WRT invariant can be written for plumbed manifolds $M(\Gamma)$ as
\begin{multline}
	\tau_{\hat{K}}^{OSp(1|2)}[M(\Gamma);\hat Q = \q]=\frac{e^{-\frac{\pi i\sigma}{4}}\,\q^{\frac{3\sigma}{4}}}{(2\hat{K}+3)^{L/2}\,(\q^{1/2}+\q^{-1/2})}
	\times\\
	{\sum_{{n_1,n_2,\ldots,n_L}\in \{1,3,\ldots,2\hat{K}+1\}}}\;\;\prod_{v\;\in\; \text{Vertices}}
	\q^{\frac{f_v(n_v^2-1)}{4}}
	\,\left(\frac{1}{\q^{n_v/2}+\q^{-n_v/2}}\right)^{\text{deg}(v)-2}
	\times \\
	\prod_{(v',v'')\;\in\;\text{Edges}}
	\Big(\q^{n_{v'}n_{v''}/2}+\q^{-n_{v'}n_{v''}/2}\Big) \Big \vert_{\q = \hat Q}~.
	\label{WRT-Gamma-0}
\end{multline}
Here again we use the Gauss reciprocity(\ref{reciprocity2}) as the summation is over odd integers to work out the steps leading to $\hat{Z}^{OSp(1|2)}[M(\Gamma);q]$. Note that, the highest integrable representation $2\hat K +1$ which fixes the $\q$ as $(2\hat K+2)$-th root of unity. However to compare the result with $OSp(1|2)$ WRT, we have to replace $\hat K +1 \rightarrow 2\hat K+3$ which is equivalent to  $\q = \hat Q$. 

Following similar steps performed for $SU(2)$, we find the following expression for $OSp(1|2)$ WRT invariant:
\begin{multline}
	\frac{1}{2\,(\q^{1/2}+\q^{-1/2})\,|\det B|^{1/2}}
	\,
	\sum_{a\in \mathrm{Coker}\,B}e^{-\pi i(2\hat{K}+3)(a,B^{-1}a)}\times\\
	\sum_{b \in 2\mathrm{Coker}\, B+\delta} e^{-\pi i\big(a,B^{-1}(b{\color{red}+BI})\big)}\lim_{q\rightarrow \hat Q} \hat{Z}_b^{OSp(1|2)}[M(\Gamma);q],
\end{multline}
where $I$ is again the column vector (\ref{eqn4.4ii}) and $\hat{Z}_b^{OSp(1|2)}[M(\Gamma);q]$ is given by the following algebraic expression:
\begin{equation}
	\hat Z_b^{OSp(1|2)}[M(\Gamma);q]\;\;=\;\;
	2^{-L} q^{-\frac{3L+\sum_v f_{v}}{4}}
	\sum_{d \;\in\; 2B\Z^L+b}F^d_1\,q^{-\frac{(d,B^{-1}d)}{4}},
	\label{Z-hat-defosp}
\end{equation}
with coefficient $F_1^d$ is obtained by following relation
\begin{multline}
	\sum_{d\;\in\;2\Z^L +\delta} F_1^d\prod_v x_v^{d_v}=\\
	\prod_{v\,\in\,\text{Vertices}}\left\{
	{\scriptsize \begin{array}{c} \text{Expansion} \\ \text{at } x\rightarrow 0 \end{array} }
	\frac{1}{(x_v+1/x_v)^{\text{deg}\,v-2}}
	+
	{\scriptsize \begin{array}{c} \text{Expansion} \\ \text{at } x\rightarrow \infty \end{array} }
	\frac{1}{(x_v+1/x_v)^{\text{deg}\,v-2}}\right\}.
	\label{eqn4.13new}
\end{multline}
Equivalently, $\hat{Z}^{OSp(1|2)}[M(\Gamma);q]$(\ref{Z-hat-defosp}) can also represented as the following contour integral:
\begin{equation}
	\hat{Z}_b^{OSp(1|2)}[M(\Gamma);q]=q^{-\frac{3L+\sum_v f_{v}}{4}}\cdot\text{v.p.}\int\limits_{|z_v|=1}
	\prod_{v\;\in\; \text{Vertices}}
	\frac{dz_v}{2\pi iz_v}\,
	\left({z_v+1/z_v}\right)^{2-\text{deg}(v)}\cdot\Theta^{-B}_b(z)~.
\end{equation}
Here $\Theta^{-B}_b(x)$ is the theta function of the lattice corresponding to minus the linking form $B$:
\begin{equation}
	\Theta^{-B}_b(x)=\sum_{d\;\in\;2B\Z^L+b}q^{-\frac{(d,B^{-1}d)}{4}}
	\prod_{i=1}^Lx_i^{d_i}
\end{equation}
and ``v.p.'' again means that we take principle value integral (\ie~take half-sum of contours $|z_v|=1\pm \epsilon$). Comparing eqns.(\ref{Z-hat-defosp},\ref{eqn4.13new}) with the $SU(2)$ expressions(\ref{F1lsu2},\ref{Z-hat-def}), we can see that the $\hat Z$ for $OSp(1|2)$ are different from $SU(2)$ q-series. We will now present explicit $q$-series for some examples.
\subsubsection{Examples}

$\bullet$ For the Poincare homology sphere(\ref{eqn3.31new}), we find the following $OSp(1|2)$ $q$-series
\begin{equation}
	\hat{Z}_{b_1}^{OSp(1|2)}=q^{-3/2}(1 + q + q^3 + q^7 + q^8 + q^{14} + q^{20} - q^{29} + q^{31} - q^{42} - 
	q^{52} +\cdots).
\end{equation}
$\bullet $ In the case of Brieskorn homology sphere(\ref{eqn3.33new}), the $OSp(1|2)$ $q$-series is  
\begin{equation}
	\hat{Z}_{b_1}^{OSp(1|2)}=q^{1/2}(1 + q + q^5 + q^{10} + q^{11} + q^{18} + q^{30} + q^{41} - q^{43} - q^{56} - 
	q^{76}\cdots). 
\end{equation}
$\bullet$ For the case of plumbing graph(\ref{eqn3.35new}), the three homological blocks are
\begin{equation}
	\hat{Z}^{OSp(1|2)}=
	\left(
	\begin{array}{c}
		1+q+q^6+q^{11}-q^{13}-q^{20}-q^{35}+O\left(q^{41}\right) \\
		q^{5/3}\left(1+q^3-q^{21}-q^{30}+O\left(q^{41}\right)\right) \\
		q^{5/3}\left(1+q^3-q^{21}-q^{30}+O\left(q^{41}\right)\right)
	\end{array}
	\right).
\end{equation}
After comparing the $q$-series for $SU(2)$ and $OSp(1|2)$, we noticed that these two $q$-series are related by a simple change of variable which is $q\longrightarrow -q$. This change of variable applies only to the series not to the overall coefficient outside the series.\\ 
$\bullet$ Lens space $L(p,q)$ is a well studied three-manifold. For $L(-5,11)\sim L(-13,29)$ whose plumbing graph is shown below, we obtain the five homological blocks \\
\begin{equation*}
	{\,\raisebox{0.0cm}{\includegraphics[width=5.0cm]{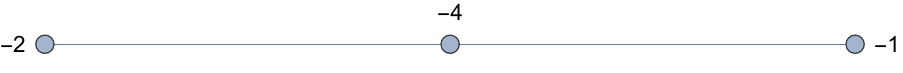}}\,}\;\;
	\stackrel{\text{Kirby}}{\sim}\;\;
	{\,\raisebox{0.0cm}{\includegraphics[width=5.0cm]{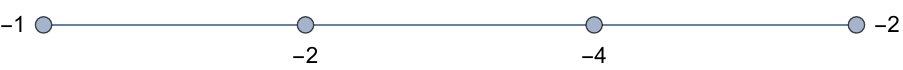}}\,}
\end{equation*}

\begin{equation}
	\hat{Z}^{OSp(1|2)}=
	\left(
	\begin{array}{c}
		q^{1/10}\\
		q^{-1/10}\\
		0\\
		q^{-1/10}\\
		q^{1/10}
	\end{array}
	\right)~~~~\text{as}~~H_1(M,\mathbb{Z})=\Z_5.
\end{equation}
$\bullet$ For the following plumbing graph, $H_1(M,\Z)=\mathbb{Z}_{13}$~,
\begin{equation*}
	\includegraphics{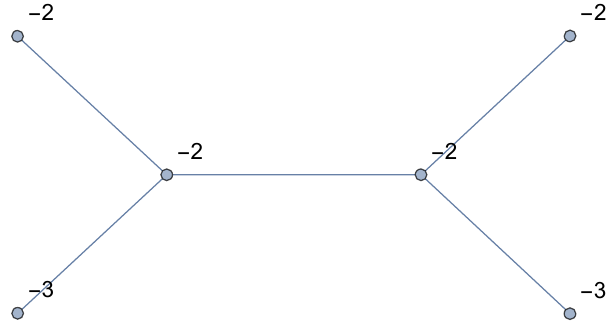}
\end{equation*}
\begin{equation}
	\hat{Z}^{OSp(1|2)}=
	\frac{1}{4}\left(\tiny
	\begin{array}{c}
		q^{-1/2}(2+2q+2q^2-2q^4-4q^5+6q^{10}-8q^{11}+4q^{13}+2q^{14}-4q^{15}+O\left(q^{18}\right)) \\
		q^{5/26}(3+2q-2q^2-4q^3-2q^7+q^8+2q^9+q^{10}-2q^{12}-4q^{13}-2q^{16}+O\left(q^{18}\right)) \\
		q^{5/26}(3+2q-2q^2-4q^3-2q^7+q^8+2q^9+q^{10}-2q^{12}-4q^{13}-2q^{16}+O\left(q^{18}\right)) \\
		q^{7/26}(4-q-2q^3-2q^4-2q^6+3q^7-2q^8-2q^{10}+q^{11}+2q^{13}-4q^{14}+2q^{15}-4q^{16}+O\left(q^{18}\right)) \\
		q^{7/26}(4-q-2q^3-2q^4-2q^6+3q^7-2q^8-2q^{10}+q^{11}+2q^{13}-4q^{14}+2q^{15}-4q^{16}+O\left(q^{18}\right)) \\
		q^{-7/26}(3+3q^2-2q^4-2q^5+4q^7-2q^8+2q^9-2q^{10}-4q^{12}+4q^{13}-4q^{14}+2q^{15}+O\left(q^{18}\right)) \\
		q^{-7/26}(3+3q^2-2q^4-2q^5+4q^7-2q^8+2q^9-2q^{10}-4q^{12}+4q^{13}-4q^{14}+2q^{15}+O\left(q^{18}\right)) \\
		q^{-11/26}(1+2q+2q^2+4q^3+3q^6-2q^7-4q^8-2q^9+2q^{11}+2q^{13}-q^{14}+2q^{16}-2q^{17}+O\left(q^{18}\right)) \\
		q^{-11/26}(1+2q+2q^2+4q^3+3q^6-2q^7-4q^8-2q^9+2q^{11}+2q^{13}-q^{14}+2q^{16}-2q^{17}+O\left(q^{18}\right)) \\
		q^{-5/26}(2+2q^2+q^3+3q^5-2q^6-2q^7-4q^8-2q^{10}+2q^{11}-2q^{12}+2q^{13}+5q^{15}-2q^{16}+2q^{17}+O\left(q^{18}\right))\\
		q^{-5/26}(2+2q^2+q^3+3q^5-2q^6-2q^7-4q^8-2q^{10}+2q^{11}-2q^{12}+2q^{13}+5q^{15}-2q^{16}+2q^{17}+O\left(q^{18}\right))\\
		q^{-15/26}(1-2q-2q^2+q^4-2q^6-2q^7-2q^8-4q^{10}-2q^{12}+2q^{13}+2q^{15}+4q^{17}+O\left(q^{18}\right)) \\
		q^{-15/26}(1-2q-2q^2+q^4-2q^6-2q^7-2q^8-4q^{10}-2q^{12}+2q^{13}+2q^{15}+4q^{17}+O\left(q^{18}\right))
	\end{array}
	\right)
\end{equation}

\vspace{0.5cm}

We have checked for many examples that under $q\rightarrow -q$ in the $OSp(1|2)$ $q$-series(not affecting the overall coefficient), we obtain the $SU(2)$ $q$-series.

\section{Conclusions and future directions}
\label{conclusion}
Our goal was to investigate $\hat Z$ for $SO(3)$ and $OSp(1|2)$ groups for negative definite plumbed three-manifolds. The change of  variable and color indeed relates invariants of framed links $\CL[\Gamma] $(\ref{so3},\ref{osp}) of $SO(3)$ and $OSp(1|2)$ to colored Jones. Such a relation allowed us to  go through  the steps of GPPV conjecture  to extract $\hat Z$ from  WRT invariants. 

Interestingly, we observe that the $\hat Z^{SO(3)}$ is same as $\hat Z^{SU(2)}$ even though the WRT invariants are different.  We know that $SU(2) /\mathbb Z_2 \equiv SO(3)$ and it is not at all obvious that the homological blocks are same for both the groups. It is important to explore other  factor groups and the corresponding $\hat  Z$ invariants. 

For the odd orthosympletic supergroup $OSp(1|2)$, we observe from our computations for many negative definite plumbing graph $\Gamma$:
$$\hat Z^{OSp(1|2)}_b(\Gamma; q) = 2^{-c} q^{\Delta_b} \left (\sum_n a_n q^n \right)$$  
whereas  their $SU(2)$  q-series is
$$\hat Z^{SU(2)}_b(\Gamma;q) = 2^{-c} q^{\Delta_b} \left (\sum_n a_n (-q)^n \right)$$

where $c\in \mathbb{Z}_+$, $\Delta_b\in \mathbb{Q}$.

The brane setup in string theory for $U(N)$ gauge group gives a natural interpretation for these $q$-series as partition function of the theory 
$T^{\mathcal G}[M]$. In principle, there should be a natural generalisation to orthogonal $SO(N)$ and symplectic group $Sp(2n)$  involving orientifolds. It will be worth investigating such a construction to obtain $\hat Z$ for $SO(N)$ group and compare with our $SO(N=3)$ results. Extension of $\hat Z$ to the two variable series for link complements\cite{Gukov:2019mnk}  is another direction to pursue. We hope to report on these aspects in future.

\begin{acknowledgements}
PR is grateful to ICTP senior associateship funding for visit where  this work was initiated with Pavel Putrov and Francesca Ferrari during summer 2019. Unfortunately due to covid, we could not pursue the collaboration through online mode. We would like to thank Pavel Putrov and Francesca Ferrari for clarifying  the notations during the initial stages. SC  would like to thank Sunghyuk Park for useful comments and discussions during the String-Math 2022 conference held at Univ. of Warsaw, Poland. SC is grateful to Dmitry Noshchenko for his comments on the manuscript. We would also like to thank Vivek Kumar Singh for his clarification on mathematica program. SC would also  like to thank the organisers of String-Math 2022 where a part of this work was presented. SC is thankful for the MHRD fellowship from IIT Bombay providing financial support to visit. PR would like to thank SERB (MATRICS) MTR/2019/000956 funding. We would like to thank the reviewers  for their valuable comments and the suggestions.

\end{acknowledgements}

\bibliographystyle{spmpsci}
\bibliography{references}

\end{document}